\documentclass[12pt]{article}
\usepackage[utf8]{inputenc}
\usepackage{amsmath}
\usepackage{graphicx}
\usepackage{adjustbox}
\usepackage{float}
\usepackage{pdfpages}
\usepackage{setspace}
\usepackage{amssymb}
\usepackage{subfigure}
\usepackage{hyperref}
\usepackage{xurl}
\usepackage{tikz}
\usepackage{caption}
\usepackage{bbm}
\usepackage{titlesec}
\usetikzlibrary{arrows.meta}
\usepackage[margin=1in]{geometry}
\usepackage{lscape}
\usepackage{pgfplots}
\usepackage{orcidlink}
\pgfplotsset{compat=1.8}
\usepgfplotslibrary{statistics}

\definecolor{mygray1}{gray}{0.9}
\definecolor{mygray2}{gray}{0.7}
\definecolor{mygray3}{gray}{0.3}
\definecolor{mygray4}{gray}{0.1}
\title{Bride Kidnapping and Informal Governance Institutions}
\author{Zachary Porreca\orcidlink{0000-0002-2213-1437}\footnote{CLEAN Unit, Via Röntgen n. 1, Milan 20136 (Italy), zachary.porreca@unibocconi.it}\\ Bocconi University}

\date{January 2024}

\begin{document}

\maketitle

\vspace{-4mm}

\begin{abstract}
	
\singlespacing
\noindent Bride kidnapping is a form of forced marriage in which a woman is taken against her will and coerced into accepting marriage with her captor. Post-Soviet Kyrgyzstan has seen a large increase in the prominence of this practice alongside a revitalization of traditional values and culture. As part of this resurgence of Kyrgyz identity and culture, the central government has formalized the authority of councils of elders called \emph{aksakals} as an arbitrator for local dispute resolution- guided by informal principles of tradition and cultural norm adherence. Bride kidnapping falls within the domain of \emph{aksakal} authority. In this study, I leverage data from a nationally representative survey and specify a latent class nested logit model of mens' marriage modality choice to analyze the impacts that \emph{aksakal} governance has on the decision to kidnap. Based on value assessment questions on the survey, men are assigned to a probability distribution over latent class membership. Utility function parameters for each potential marriage modality are estimated for each latent class of men. Results suggest that living under \emph{aksakal} governance makes men 9\% more likely to obtain a wife through bride capture, with men substituting kidnapping for choice marriage modalities such as elopement and standard love marriages.

\end{abstract}

\textbf{Keywords}: Bride Kidnapping; Forced Marriage; Informal Institutions; Kyrgyzstan 

 \textbf{JEL Codes}: J12, K42, N35, P37, O17, J16, Z10
 
 \clearpage
 
\doublespacing
\vspace{-20mm}

\pagebreak
\section{Introduction}
Violence against women is unfortunately a global phenomenon. Traditional forms of gendered violence can provide insights into the dynamics of similar phenomena in contemporary Western societies. Few forms of such violence are as stark as that of bride capture. This is the practice of men forcibly kidnapping women against their will, in the hopes of coercing, through violence and social stigma, her into marriage (Sheilds 2006). Marriages stemming from capture predictably are associated with adverse outcomes for the women involved. Steiner \& Becker (2019) document that these marriages result in couples that are less matched on personality traits. Sheilds (2006) documents that these kidnapped women often become victims of rape and domestic violence. Becker et al. (2017) demonstrate that these marriages are associated with lower infant birth weight. Kidnapped women are less likely to be engaged in the labor force (Arabsheibani et al. 2021) and increases in kidnap risk lead to less educational investment for young women (Bazarkulova \& Compton 2021).

Less is known of the dynamics regarding the risk of bride capture and its interplay with traditional value systems. Bride capture is most well documented and most prevalent in the Central Asian country of Kyrgyzstan. Despite commonly held beliefs, non-consensual bride kidnapping was not common to Kyrgyzstan prior to the country's Soviet era (Kleinback \& Salimjanova 2007). Nedoluzhko \& Agadjanian (2015) show that rates of kidnap marriage are declining from their peak in the late 1990's. However, some estimate that still up to 30\% of marriages in the country are the result of bride capture (Sheilds 2006; Werner et al. 2018). Despite the practice being formally illegal, very few perpetrators are ever brought to justice (Sheilds 2006). One understudied reason for this unfortunate reality is that kidnapping cases are often not brought before modern criminal courts, but rather are placed before traditional elder courts called \emph{aksakals} (Sheilds 2006; Aktas 2021). These \emph{aksakal} courts are prone to view bride capture as carrying on an important cultural legacy for the country, and as such leave the victims without legal remedy or protection (Beyer 2006; Aktas 2021).

While we know much about the history, context, and dynamics of bride capture, no work has examined its drivers. Here, I shift the focus towards the men choosing whether to engage in this practice and provide quantitative evidence of the determinants of their decision to kidnap. The lack of research attention to the risk factors behind bride capture was noted in Werner et al. (2018). In particular, I focus on the role that traditional governance institutions play in mens' choice to pursue marriage through non-consensual means\footnote{Kleinbach et al. (2005) note that future research would do well to analyze the impacts of \emph{aksakal} courts on bride capture incidence.}. Bride capture is a crime, yet its continued prevalence in the country remains until this day. Becker's (1968) foundational model of criminality sees criminals weighing their probability of punishment, potential sanctions, and rewards if successful when deciding to engage in crime. In Kyrgyzstan institutional variation plays an important role in determining the likelihood of punishment kidnappers face and undoubtedly plays an important role in the calculus of a kidnapper's decision making. By turning our focus towards the perpetrators of this crime, we can begin to think about how we can optimally address a practice that exists as both a violation of human rights and a hindrance to economic development. 

A lack of detailed longitudinal data and appropriate settings for causal identification have precluded much quantitative work on this topic\footnote{Kleinbach \& Babayarova (2013) provides a descriptive analysis of a program that included an informational treatment, providing context on kidnappings to young Kyrgyz men and a ``pledge" to refuse to marry a kidnapper for Kyrgyz women, with the intention of reducing bride capture incidence. Reductions in kidnap incidence over the following year were observed for the villages the researchers visited. However, the research design was non-causal. }. I circumvent these obstacles by exploiting a rich survey dataset and specifying a structural model of discrete choice to analyze the determinants of individual Kyrgyz men's decisions of marriage modality. This model, utilizing a wealth of value assessment questions, classifies men as either being traditionally minded or more contemporary in their value system. Then, for each class of men estimates the parameters of a utility maximization model to provide insights into the trade offs and choices inherent in a man's marriage decisions. In particular, this model is able to quantify the impact that \emph{aksakal} governance, rather than state institutional governance, has on bride capture. 

The remainder of the paper is organized as follows. Section II provides institutional details regarding bride capture, marriage broadly, and \emph{aksakal} governance in Kyrgyzstan. Section III analyzes survey responses from kidnappers to provide descriptive evidence of the rationales behind the decision to kidnap. Section IV describes the `Life in Kyrgyzstan' survey data employed here. Section V describes the structural model of Kyrgyz men's marriage modality choices and its estimation. Section VI presents the results from estimating this model and Section VII concludes. 
 
\section{Marriage and Informal Governance in Kyrgyzstan}
``\emph{Ala kachuu}" in modern usage is the Kyrgyz term for the non-consensual kidnapping of a woman for marriage. Kleinbach et al. (2005) and Kleinbach \& Salimjanova (2007) provide a detailed history and social context for the practice. In short, in the pre-Soviet era most marriages were arranged and necessitated the payment of a ``kalym" or bride payment by the groom's family. \emph{Ala kachuu} existed then as form of consensual elopement (mock kidnapping) that largely served to avoid potential familial disapproval or the expenses of arranged marriages (Kleinbach \& Salimjanova 2007). In the same era, non-consensual bride capture was nearly exclusively conducted by wealthy and powerful families due to the risk of tribal conflict that such incidents could spur (Kleinbach \& Salimjanova 2007). The Soviet era brought reform to the country, outlawing non-consensual marriages (including arranged marriages and kalym payments)  and introduced legislation mandating equality for men and women under the law (Kleinbach \& Salimjanova 2007). Kleinbach et al. (2005) documents that in this period \emph{ala kachuu} came to represent a new form of elopement for couples without parental consent, and that this form of the practice eventually evolved into the non-consensual variety observed in more recent years.\footnote{In some ways, bride kidnapping is analogous to the wife sales that took place in Victorian England and are detailed in Leeson et al. (2014). In that context, men found a unique institutional means to exit marriages that they did not value highly, at a lower cost than divorce. Here, often low-value men find an institutional arrangement that allows them to obtain brides of ``higher value" than they would be able to otherwise obtain through more normal means.} 

Following the fall of the Soviet Union, the country has seen a resurgence of traditional beliefs and practices- with a particularly negative impact for the status of women (Heyat 2004). The end of the Soviet era and the first years of independent Kyrgyzstan saw the highest rates of forced non-consensual bride kidnapping, with estimates from Nedoluzhko \& Agadjanian (2015) placing up to 64\% of ethnic Kyrgyz marriages in the 1980's and over 55\% from the first 15 years of independence as originating in the practice. 

The emergence of an independent Kyrgyzstan and the accompanying cultural movement towards a rebirth of ``traditional" practices and culture led to the reemergence of the \emph{aksakal} courts as local institutions. The \emph{aksakals}, or ``white beard" community elders, were given legal authority in the country's first constitution in 1993 (Beyer 2006). The traditional role of \emph{aksakals} in Kyrgyz culture is argued to date back to ancient times (based upon mention in the nation's epic poem \emph{Manas}) where the \emph{aksakal} would resolve disputes between nomadic clans (Atkas 2021). These courts of local elders were first formally established in the rural countryside, where the ``authority of elders" and ``force of customary law" was believed to remain since pre-Soviet times (Beyer 2015). Early on, men who already held positions of prominence in their village, often men who had previously held important roles in Soviet collective farms, were appointed as official \emph{aksakal} judges. However, as the country experienced it's resurgence in traditional beliefs, practices, and religion, men returning from Islamic pilgrimage to Mecca (the \emph{hajj}) would often be appointed to the courts (Heyat 2004). 

The \emph{aksakal} courts are intended as a means of local dispute resolution and by law are charged to “judge according to moral norms that reflect the customs and traditions of the Kyrgyz” (Art. 1, I, 2, Law on the Aksakal Courts, June 2002). These courts do not pass judgment on the basis of written or formalized law. Instead, decisions are rendered based upon the whims of the elders who consider traditional cultural norms\footnote{In some regards, including governance based in part upon traditional beliefs, \emph{aksakal} courts are similar to the policing function of Native American military societies detailed in Crepelle et al. (2022).}. The \emph{aksakal} courts are an informal institution in that, despite being legally recognized, their internal procedures are not formalized and by official law individuals involved in disputes must agree to have their case be decided upon by the \emph{aksakal} (Beyer 2006)\footnote{The \emph{aksakals} are also defined as informal institutions throughout the literature on the topic. A lengthy discussion of this distinction is present in Sheranova (2020).}. However, where the weight of tradition is heavy these courts often become the de-facto arbitrator of disputes. A norm in Kyrgyz culture that dictates that outsiders should not be brought into problems in which they or their relatives are not involved often precludes parties to a dispute from taking up their case in formal state courts (Atkas 2021).  The \emph{askakals} have providence over disputes regarding land and water, livestock theft, family law, domestic violence, and some small crimes (Beyer 2015). Punishments rendered by these courts can include corporal punishment (such as stoning or whipping) and detention in technically illegitimate facilities (Handrahan 2004).

Because domestic violence is within the \emph{aksakals'} domain, police typically refer cases of bride capture there rather than to official state courts (Sheilds 2006). This is unfortunate, as these courts have been known to punish the kidnapped women for the offense of trying to leave the non-consensual union (Bagirova 2021). The \emph{askakal} tend to focus on potential reconciliation among parties\footnote{The emphasis on avoiding conflict between parties in favor of fostering reconciliation is similar to that documented by Thompson (2023) in his work on the La Cosa Nostra mafia's dispute resolution mechanisms in America.} (Shields 2006) and due to their focus on tradition, their judgments often act to uphold patriarchal societal norms (Handrahan 2004). These courts tend to view the bride capture as a legitimate means of marriage (Atkas 2021). Unsurprisingly, all of this leads to very few cases of bride kidnapping ever being brought before the police or formal state courts (Sheilds 2006). There is almost no risk of punishment for kidnappers.

In the framework of Becker's (1968) foundational model of criminal decisions, potential criminals choose to commit the crime if its expected benefits outweigh its expected costs. In the context of kidnappings, Detotto et al. (2015) assumes a model based on the same foundation and provides evidence that would-be kidnappers are weighing incentives and acting ``rationally" in their decision making process.\footnote{Detotto et al. (2014) takes a similar approach, but instead takes the kidnapping as ``given" and models the duration for which a victim would optimally be held for ransom.} In the Kyrgyz context, rational actors observing their community environment, and perceiving the lowered risk of punishment associated with \emph{aksakal} governance would require a lower threshold for a potential gain in order to elect to engage in bride capture. This means that, when governed by an \emph{aksakal} rather than the state, the requisite relative benefits of kidnapping that would drive a potential kidnapper to make the decision are lessened. Survey data for these factors are included in the subsequent section, but it is important to note that when governed by traditional elder courts \emph{how much} of a reduction in these factors is needed to push individuals towards bride capture is theoretically reduced. \emph{Aksakal} governance should increase incidents of kidnapping among rational men deciding on marriage modality. 

Despite \emph{aksakal} courts existing throughout the entirety of Kyrgyzstan, their influence is much greater in rural villages (Atkas 2021). In fact, in their urban form the \emph{aksakals} are often not Kyrgyz and can be women themselves (Beyer 2015). In this non-traditional role, their decisions are not rendered based upon Kyrgyz tradition and customs, but simply by the judgment of the elder. However, as the official state government has a much larger presence in urban areas, the influence and power of the \emph{aksakals} is greatest in the rural villages where a sentiment exists that the state does not extend (Atkas 2021). For the empirical section of the study at hand, I rely upon the community section of the survey for determining which communities are governed by \emph{aksakal}. Localities are classified as being governed by \emph{aksakals} based upon how the survey respondent (in this case a local with official state authority) answered the prompt ``How is the decision normally made at this kind of meeting?". This question is asked following a series of questions about local meetings to resolve disputes and manage local public good provision. Localities are identified as being governed by the \emph{aksakal} if the survey respondent answered that question with ``Community leaders, eg. \emph{aksakals} make a decision,and other community members accept it". Of the 111 unique communities in my final analysis data, 23 are governed by \emph{aksakal}.

\section{Why Do Men Choose to Kidnap?}
Included in the survey data that my empirical analysis relies upon are a series of questions regarding the ``possible advantages" of bride capture. These questions are framed as ``In general, how would you assess possible advantages of marriage through bride kidnapping?", followed by a series of statements that the respondent is instructed to rate the relative importance of from 1 (``the most important) to 4 (``not important at all"). Here, I provide a descriptive analysis of the responses to this question, limited to a subset of male respondents who have either been married through bride capture (including so called ``mock" kidnappings) or in another survey question responded that they had previously taken part in an unsuccessful bride kidnapping as a potential husband (~$n=984$). Examining the supposed benefits of this practice as perceived by those who have chosen to engage in it allows for an understanding of the drivers of this practice. These responses motivate the inclusion of variables in the utility functions specified in a later section. These statements are listed in the table below.

\begin{table}[H]
	\centering
	\resizebox{.8\textwidth}{!}{%
\begin{tabular}{c|c}
	\hline
	\multicolumn{2}{c}{\textbf{Statement}} \\
	\hline 
	\hline
	Q1 & Reduces wedding expenses \\
	Q2 & Reduces kalym \\
	Q3 & It is a way to avoid delayed marriage or prolonged search for spouse \\
	Q4 & It is a way to avoid lengthy negotiations between couple's parents \\
	Q5 & It is carrying on social custom \\
	Q6 & It is a way to avoid disapproval of a marriage by couples' parents \\
	\hline
\end{tabular}
}
\caption{Table depicting statements survey respondents were asked to rate the importance of regarding ``possible advantages" of bride capture.}
\label{}
\end{table}

The proportions of self-admitted kidnappers rating each of these statements as ``quite important or "the most important" are depicted in the figure below. 

\medspace

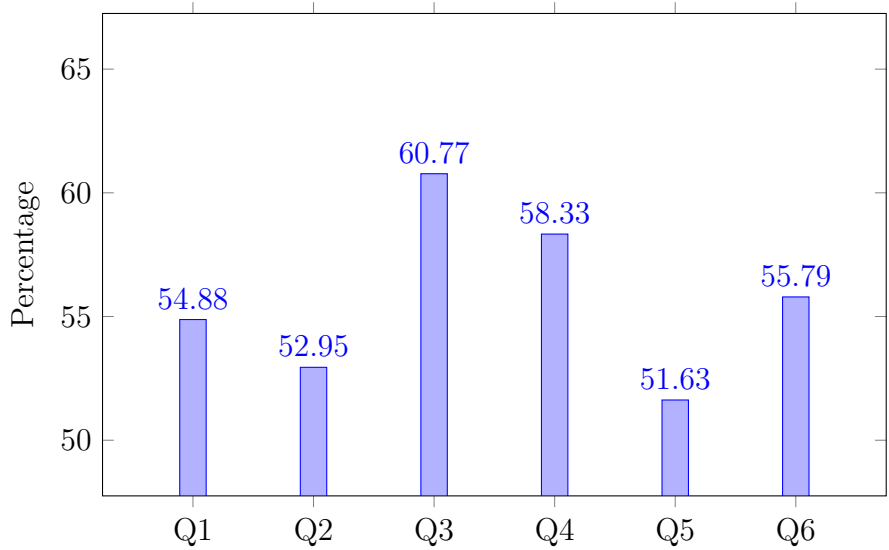
\begin{figure}[H]
	\begin{centering}
\begin{tikzpicture}
\begin{axis}[
width=12cm,    
height=8cm,
ybar,
enlargelimits=0.15,
legend style={at={(0.5,-0.20)},
	anchor=north,legend columns=-1},
ylabel={Percentage},
symbolic x coords={Q1, Q2, Q3, Q4, Q5, Q6},
xtick=data,
nodes near coords,
nodes near coords align={vertical},
every node near coord/.append style={font=\normalsize}, 
ymin=50, ymax=65,
]
\addplot coordinates {
	(Q1,54.878) 
	(Q2,52.947) 
	(Q3,60.772) 
	(Q4,58.333) 
	(Q5,51.626) 
	(Q6,55.792)};
\end{axis}
\end{tikzpicture}
	\caption{Figure depicting the proportion of self-admitted kidnappers answering affirmatively to survey questions relating to the benefits of kidnapping.}
	\label{}
		\end{centering}
\end{figure}

First, it is clear that each of these ``advantages" was considered as important to those who choose to kidnap. The largest proportion of kidnappers seem to consider the expediency of kidnapping as a major advantage, both in avoiding the effort required by the search of a spouse and in the necessary negotiations with a prospective spouse's family. Survey responses regarding negotiations between parents and avoidance of parental disapproval point to the importance of familial and individual social status. Werner et al. (2018) note's that men who kidnap have often been rejected in the past. Leeson \& Suarez (2017) similarly show that husbands of child brides are of lower quality and less desirable, namely that they are of lower castes and less education. The importance of reducing kalym (bride payments) and wedding expenses relates to the importance of household wealth in the decision to kidnap. The role that poverty and the costs associated with consensual marriages have in perpetuating bride capture was also noted in Werner et al (2018). Lastly, despite being the relative least important factor, more than half of this set of respondents did point to the importance of bride capture as a social custom. This motivates the importance of the latent class framework employed in the model, aimed at separating those individuals with traditional value systems from those of more modern value sets. 

\section{Data}
The data employed comes from The ‘Life in Kyrgyzstan’ Study, a longitudinal survey of households, individuals, and communities in Kyrgyzstan.\footnote{This data is made available by application to IZA.} The first wave of this study was conducted in 2010 with its most recent wave conducted in 2019. Here, I rely primarily on the 2016 wave of this study, as it was the first wave to question men about marriage background with detailed categorization of their marriage modality. Given the relative lack of official statistics for Kyrgyzstan, this survey has become the dominant source of in-depth information about life within the country. Further, it is potentially the largest data source with information about the phenomenon of bride kidnapping within the country. It has been used in several foundation studies on the topic including Becker \& Steiner (2017), Werner et al. (2018), Steiner \& Becker (2019), Arabsheibani et al. (2021), and Bazarkulova \& Compton (2021),. 

This survey attempts  to return to the same households and individuals with each wave, observing the life course of the country's population. To construct the sample I utilize here, I limit the sample to men over the age of 18 (the mean age for mens' first weddings in Kyrgyzstan has consistently been in the low 20's (Nedoluzhko \& Agadjanian 2015)) whose responses were labeled by the surveyor as ``reliable" and for who there was no missing data. This results in a sample of 3340 individual observations. Individual data is drawn from the individual portions of the surveys, details about the entirety of their households from the household portion, and details about communities (unsurprisingly) from the community section. Average community ``kalym" (bride price or reverse dowry) prices were constructed as the average payments (in units ``sheep that could be bought with the sum paid") for all unique marriages documented in the 2011 and 2012 survey waves, as these were the only two waves to include this question. More details regarding the construction of the variables used in this analysis are included alongside the utility functions in which they are introduced. 

\section{Modeling and Estimation}
Here, I outline my model and estimation strategy. Given the richness of the survey data available, I adopt a latent class framework based upon that employed in Greene and Hensher (2003). Latent class modeling allows for heterogeneity across the model's agents, so that not all individuals need to respond to incentives in the same manner. The combination of the latent class framework with the textbook nested logit formulation of Greene (2018) is similar to that employed in Wen et al. (2012), who analyze heterogeneous preferences in public transit demand.

By assumption there are two classes of men in our sample: those with traditional value systems and those of a more contemporary leaning. Each of these two classes is posited to have separate utility function parameters. Class membership is predicted based upon survey responses related to the individual's value system. This is detailed in greater depth in the next section. 

Men in Kyrgyzstan face the following choice structure in selecting the modality by which they will obtain a bride. They can forgo marriage entirely, have their family arrange a marriage, kidnap a bride against her will, or pursue a ``choice" marriage.\footnote{The terminology of ``choice" marriage to describe matches that are not derived from kidnapping or arrangement is adopted from Nedoluzhko \& Agadjanian (2015).} Within the choice marriage nest, men can select a ``traditional" love marriage with a willing partner or can engage in a ``mock" consensual kidnapping, a form of elopement (Nedoluzhko \& Agadjanian 2015). This choice structure is illustrated below:

\begin{center}
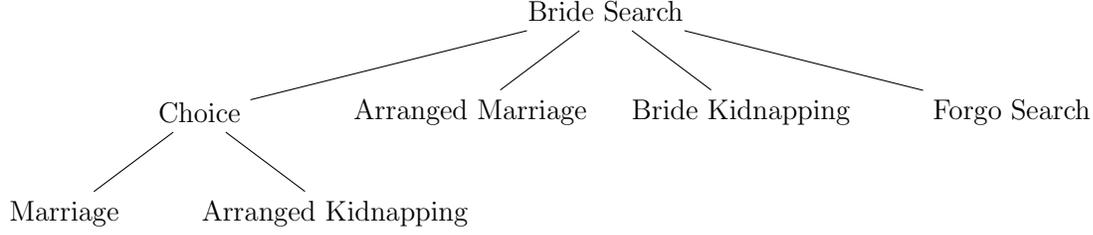
\begin{figure}[H]
\begin{tikzpicture}[scale=0.9, every node/.style={scale=0.9}, level 1/.style={sibling distance=40mm},
                   level 2/.style={sibling distance=40mm}]

  \node {Bride Search}
    child { node {Choice }
      child { node {Marriage} }
      child { node {Arranged Kidnapping} }
    }
    child { node {Arranged Marriage}
    }
    child { node {Bride Kidnapping}
    }
    child { node {Forgo Search}
    };
\end{tikzpicture}
\caption{Diagrammatic representation of modeled choice structure of Kyrgyz men's marriage modality decision.}
\label{}
	\end{figure}
\end{center}

Individuals, observing the state of their local area, individual circumstances, and with an inherent set of values (which determine class membership), select a modality by which to obtain a bride. Individual ~$i$, belonging to class ~$c$, resident in location ~$j$, choosing a modality ~$m$ has a utility function of the basic form:

\begin{equation}
    U_{ijm|c}=\beta_m^c X_{ijm}+\epsilon_{ijm|c}
\end{equation}

The probability that an individual selects a modality ~$m$ is equal to:

\begin{equation}
   P_{ij}(m)= \sum_c \bigg(P_{ij}(m|n,c)\cdot P_{ij}(n|c)  \cdot H_{ij}(c)\bigg)
\end{equation}

Here, ~$n$ indexes the nests (choice marriage, arranged marriage, bride kidnapping, and forgone search) in which lower hierarchical choices exist. It is worth stating that in all but the choice marriage nest the ~$P_{ij}(m|n,c)$ term is equal to 1, since there are no lower hierarchical decisions within those nests.

Here, ~$H_{ij}(c)$ represents the probability that individual ~$i$ belongs to class ~$c$. I adopt the multinomial logit form for these class membership probabilities as in Green (2001; 2003). This is expressed as:

\begin{equation}
    H_{ij}(c)=\frac{exp(z_{i}\theta_c)}{\sum_c exp(z_{i}\theta_c)}
\end{equation}

Here, the ~$z_i$ terms represent a set of observable characteristics that are predictive of class membership. In this context, these characteristics are drawn from individual survey responses that point to the values and beliefs of the individual. 

As mentioned earlier, for all ``nests", ~$n$, with the exception of the choice marriage nest, ~$(P_{ij}(m|n,c)\cdot P_{ij}(n|c))=P_{ij}(n|c)$. Only one choice exists in each of those nests. However, within the choice marriage nest:

\begin{equation}
    P_{ij}(m|n,c)=\frac{exp(\beta_m^c X_{ijm|n})}{\sum_{m \in n}exp(\beta_m^c X_{ijm|n})}
\end{equation}

and 

\begin{equation}
    P_{ij}(n|c)=\frac{exp(\beta_n^c X_{ijn}+\Gamma_{in}^c)}{\sum_n exp(\beta_n^c X_{ijn}+\Gamma_{in}^c)}
\end{equation}

Where ~$\Gamma_{in}$ is the inclusive value of the ~$n$th nest, defined as:

\begin{equation}
    \Gamma_{in}^c=ln \bigg( \sum_{m\in n} exp(\beta_m^c X_{ijm|n}) \bigg)
\end{equation}

This is the common form of nested logit probabilities provided in Greene (2018). For all of the other modalities/nests, the inclusive value is simply equal to that modality's utility. For the other nests without branches, where the nest is equivalent to the modality choice, these probabilities take the common mulitnomial logit probability form of McFadden (1978), in which each choice has a separate set of parameter values, with the addition of the inclusive values into the probability formulation as in equation (5).

Thus, combining this with our latent class probabilities, for all modalities outside of the choice marriage nest, the probabilities are expressed as:

\begin{equation}
    P_{ij}(n)= P_{ij}(m)= \sum_c \bigg(\frac{exp(\beta_n^c X_{ijn}+\Gamma_{in}^c)}{\sum_n exp(\beta_n^c X_{ijn}+\Gamma_{in}^c)}  \cdot \frac{exp(z_{i}\theta_c)}{\sum_c exp(z_{i}\theta_c)}\bigg)
\end{equation}

And for the choice marriage modalities, the probabilities of choosing a given modality are equal to:

\begin{equation}
         P_{ij}(m)= \sum_c \bigg(\frac{exp(\beta_n^c X_{ijn}+\Gamma_{in}^c)}{\sum_n exp(\beta_n^c X_{ijn}+\Gamma_{in}^c)} \cdot \frac{exp(\beta_m^c X_{ij})}{\sum_{m \in n} exp(\beta_m^c X_{ijm})}  \cdot \frac{exp(z_{i}\theta_c)}{\sum_c exp(z_{i}\theta_c)} \bigg)
\end{equation}

I proceed with one-step full information maximum likelihood estimation (FIML). The log-likelihood function can be defined as follows with an indicator variable ~$y_{ijm}$ introduced to represent each individual's choice (as discussed in Cameron and Trivedi (1995)):

\begin{equation}
     \mathcal{LL}()=\sum_i\sum_m  \bigg(y_{ijm} \cdot ln(P_{ij}(m)) \bigg)
\end{equation}

Now, with the basic formulation of the model outlined, I can begin specifying the utility functions (beyond the basic form employed above) which will populate this model. For each modality, the true utility function is an expected utility function relying on many characteristics that remain unobservable to the researcher. As such, I approximate these expected utility functions linearly with observable proxy variables replacing the unobservable variables that would enter the true expected utility function.

First, I assume that the utility derived from the outside option (forgoing the search for a partner) can be normalized to zero for all classes of individuals:

\begin{equation}
    U_{ij(m=Forgo)|c}=0
\end{equation}

Next, for individuals choosing to kidnap a bride, the utility that they expect to derive from this decision is a function of the risk of punishment they face (proxied by their community's being governed by an \emph{aksakal} court and police presence within their community), average local kalym prices, their ``marketability" in finding a bride through more standard means (proxied by employment status, household income and household assets), and the social standing of their household (proxied by informal loan availability to members of their household and their household's being an event host). Here, second home ownership and vehicle ownership are used to account for household assets since only 8\% of households own a second home (versus 96\% that own a first home) and only 44\% of households own a car or truck. Only 32\% of households host events in a given year, making this a good proxy for a household's status within the community. Further, Aldashev (2023) argues that in the Kyrgyz context spending on events and festivities is primarily a means to maintain informal networks. Lastly, in this country informal loan availability serves as a symbol of both status within the community and social capital (Urbaeva et al. 2018). The ~$\epsilon$ terms present throughout all of these utility functions represent individual heterogeneity. I approximate the true expected utility function linearly as:

\begin{equation}
\begin{split}
U_{ij(m=Kidnapping)|c}&= \beta^c_0+\beta^c_1(Aksakal\;Governance)_j+\beta_2^c(Police\;Presense)_j+\\& +\beta_3^c(Kalym\;Price)_j+\beta_4^c(Household\;Income)_i+ \beta_5^c(Second\;Home\;Ownership)_i\\& +\beta_6^c(Vehicle\;Ownership)_i+\beta_7^c(Loan\;Availability)_i+\beta_8^c(Event\;Host)_i\\&+\beta^c_9(Employed)_i +\epsilon_{ij(m=Kidnapping)|c}
\end{split}
\end{equation}

Individuals finding a bride through an arranged marriage orchestrated by their family derive utility in a similar manner, but given the cultural and formal legal acceptability of the practice do not face any risk of punishment. An approximation of their expected utility function can be represented as:

\begin{equation}
\begin{split}
U_{ij(m=Arranged)|c}&= \alpha^c_0+\alpha_1^c(Kalym\;Price)_j+\alpha_2^c(Household\;Income)_i\\&+ \alpha_3^c(Second\;Home\;Ownership)_i +\alpha_4^c(Vehicle\;Ownership)_i\\&+\alpha_5^c(Loan\;Availability)_i +\alpha^c_6(Event\;Host)_i+\alpha^c_7(Employed)_i\epsilon_{ij(m=Arranged)|c}
\end{split}
\end{equation}  

For couples agreeing to a kidnapping (elopement or ``mock" kidnapping), the expected utility function is hypothesized to be largely similar to that of individuals engaging in non-consensual kidnappings. However, here the risk of punishment is much less. As such, the police station proxy for punishment risk is left out of the utility function specification. However, the \emph{aksakal} governance variable remains, as these courts are likely to sanction (and even encourage) such matches. Variables representing individuals' marriage ``marketability" and household social standing remain, as status remains important in obtaining a bride through choice marriage modalities. Shields (2006) argues that many men who are deemed ``socially undesirable" due to criminal background, disability, or other characteristics face greater difficulty in convincing a woman or her family to consent to a marriage and as such may resort to kidnapping as an alternative. This is also taken to be the utility function of the overall choice marriage nest (albeit with different parameter values).

\begin{equation}
\begin{split}
U_{ij(m=Arranged\;Capture)|c}&= \lambda^c_0+\lambda^c_1(Aksakal\;Governance)_j + \lambda_2^c(Kalym\;Price)_j\\&+\lambda_3^c(Household\;Income)_i + \lambda_4^c(Second\;Home\;Ownership)_i\\&+\lambda_5^c(Vehicle\;Ownership)_i+\lambda_6^c(Loan\;Availability)_i+\lambda_7^c(Event\;Host)_i \\&+\lambda^c_8(Employed)_i+\epsilon_{ij(m=Arranged\;Capture)|c}
\end{split}
\end{equation}

Lastly for those grooms electing for a choice, non-elopement, marriage,  \emph{aksakal} governance is largely irrelevant and I specify their expected utility function to solely be a function of their and their household's social status, and local kalym prices. 

\begin{equation}
\begin{split}
U_{ij(m=Love\;Match)|c}&= \psi^c_0+ \psi_1^c(Kalym\;Price)_j+\psi_2^c(Household\;Income)_i \\&+ \psi_3^c(Second\;Home\;Ownership)_i+\psi_4^c(Vehicle\;Ownership)_i\\&+\psi_5^c(Loan\;Availability)_i+\psi_6^c(Event\;Host)_i\\&+\psi^c_7(Employed)_i +\epsilon_{ij(m=Love\;Match)|c}
\end{split}
\end{equation}

It bares noting that with the model specified as such the same variables enter multiple utility functions with different parameters. This approach is intentional. It allows for heterogeneous impacts of the same underlying environmental and individual characteristic variables across the probability of making different choices. For example, there is no reason to expect the variables proxying the probability of punishment to impact the choice to pursue a completely legal marriage option in the same way it impacts the choice to pursue an illegal path. This heterogeneity and the associated complexity of the model comes at the cost of computational difficulty. As such, the model is estimated through full information maximum likelihood methods, employing the Simulated Annealing algorithm. This algorithm is able to avoid converging to local optima and is well suited for finding the global optimum in complex parameter spaces such as that of our high dimensional model (Goffe et al. 1994). It is able to provide consistent solutions that are largely independent of initial starting values. (Goffe et al. 1994). Further, this algorithm (under certain conditions) guarantees eventual convergence, albeit at the expense of computational time (Henderson et al. 2003). Uncertainty quantification of parameter values using bootstrapping or other methods of random sample inference are infeasible given the lengthy computational time necessitated for convergence to the global optimum, and as such is conducted as in the same manner as Amilon (2003) and Porreca (2024), in which following convergence to a global optimum using a heuristic-solver optimal values are used to compute the gradient/Hessian computationally to then estimate the covariance matrix. Here, I employ the robust estimator of the covariance matrix from Greene (2018).

\begin{equation}
\hat{Avar}(\hat{\beta})\approx\bigg[-\bold{H}(\hat{\beta})\bigg]^{-1}\bigg[\bold{g}(\hat{\beta})\bold{g}(\hat{\beta})^T\bigg]\bigg[-\bold{H}(\hat{\beta})\bigg]^{-1}
\end{equation}   

 In the above equation the ~$\bold{g()}$ terms represent the gradients of parameter estimates, while the ~$\bold{H()}$ represents the Hessian matrix for parameter estimates. In the next section, I detail the survey questions used for estimating latent class membership.
 
\subsection{Estimating Latent Classes}
Latent class estimation allows for heterogeneity among the agents of the model. There are differences among individuals, and as such the parameters that have been specified to govern the utility functions of these men are certainly not uniform. By estimating a probability distribution over the class membership of each individual, I am able to estimate two separate sets of parameter values for each of the utility functions described. These classes are meant to capture variation in the value sets of Kyrgyz men. While there is in reality undoubtedly much more individual specific heterogeneity than this model allows, estimating average parameter values for two broad classes of individuals: the traditional/conservative and those of more contemporary value sets, allows us to at least somewhat unpack this heterogeneity. 

My goal is to assign men to one of two latent classes, which may impact the decision they make in regards to how to obtain a bride. To an extent, I build on the approach of Bazarkulova \& Compton (2021), in approximating an index of traditional values. Here, however, I do this at the individual level (rather than household) in an attempt to classify the individual making the marriage decision. The survey information that I employ in building the class membership probabilities are primarily from two sections of the individual portion of the `Life in Kyrgyzstan Survey' titled ``Values and Gender Attitudes" and ``Marriage Practices". For the ``Values and Gender Attitudes" portion, I code dummy variables for survey respondents indicating that they ``strongly agree" with each of several statements regarding the role of women in marriage (response values of 3 and 4). For the``Marriage Practices" portion, I code dummy variables for survey respondents indicating the relevant ``characteristic of a spouse" is ``the most important" or ``quite important" (response values 1 and 2). The exact survey questions utilized are listed in the table below. In addition to these, I create an indicator variable for if the man is either Kyrgyz or Kazakh, as Bazarkulova \& Compton (2021) discuss these two ethnic groups are the only two in the country that widely engage in the practice of bride capture, as well as indicator built from the community section of the survey for whether the municipality of the man's residence is a village (as opposed to a city). 

\begin{table}[H]
	\centering
	\resizebox{\textwidth}{!}{%
		\begin{tabular}{c|c}
	\hline
	\multicolumn{2}{c}{\textbf{Values and Gender Attitudes}} \\
	\hline
	Statement & Response \\
	\hline
	Important decisions should be made by the husband rather than the wife. & Strongly Agree (3 or 4) \\
	A man’s job is to earn money; a woman’s job is to look after the home
	and family. & Strongly Agree (3 or 4) \\
	A woman is really fulfilled only when she becomes a mother. & Strongly Agree (3 or 4) \\
	A husband’s career should be more important to the wife than her own. & Strongly Agree (3 or 4) \\
	A university education is more important for a boy than for a girl. & Strongly Agree (3 or 4) \\
	Woman should not work outside her home due to religious
	considerations & Strongly Agree (3 or 4) \\
	Being a housewife is just as fulfilling as working for pay. & Strongly Agree (3 or 4) \\
	A working woman can establish just as warm and secure relationship with her children as a mother who does not work. & Strongly Agree (3 or 4)\\
	Both the husband and the wife should contribute to the household income. & Strongly Agree (3 or 4) \\
	\hline
	\multicolumn{2}{c}{\textbf{Marriage Practices}} \\
	\multicolumn{2}{c}{Irrespective of whether you are married or not, how important are following characteristics of spouse?} \\
	\hline
	Characteristic & Response \\
	\hline
	Respectful, doesn't complain and criticize & ``The Most Important" or``Quite Important" (1 or 2) \\
	Obedient & ``The Most Important" or ``Quite Important" (1 or 2) \\
	Confident & ``The Most Important" or ``Quite Important" (1 or 2) \\
	Intelligent & ``The Most Important" or ``Quite Important" (1 or 2) \\
	Respected by Society & ``The Most Important" or ``Quite Important" (1 or 2) \\
	\hline
	\multicolumn{2}{c}{\textbf{Trust}} \\
	\multicolumn{2}{c}{How much do you agree with the following statements?} \\
	\hline
	Statement & Response \\
	\hline
	In general, you can trust people. & Strongly Agree (3 or 4) \\
	Nowadays, you cannot rely on anybody. & Strongly Agree (3 or 4) \\
	In this community, you have to be cautious, otherwise someone is likely to take advantage of you. & Strongly Agree (3 or 4) \\
		\hline
	\multicolumn{2}{c}{\textbf{Trust}} \\
	\multicolumn{2}{c}{How much do you generally trust the following?} \\
	\hline
	Group & Response \\
	\hline
	Family Members & Trust (3 or 4) \\
	Neighbors & Trust (3 or 4) \\
	People who you do not know & Trust (3 or 4) \\
	People from other ethnic or linguistic groups & Trust (3 or 4) \\

\end{tabular}
}
\caption{Table of survey responses coded as indicator variables for latent class estimation.}
\label{}
\end{table}

To force these two classes to capture differences between ``traditional" and more modern value sets, we restrict parameters for the ``values and gender attitudes" and ``marriage practices" sections of the survey to opposing negatives and positives. Class 1 is restricted to positive coefficients for agreement with the more patriarchal sentiments (including the first three characteristics in the ``marriage practices" section of the above table) and negative coefficients for agreement with the equality espousing statements (including the last two statements in the ``marriage practices" section of the table). The opposite restrictions are made for class 2. These restrictions ensure that espousing traditional values increases the probability over a man's assignment to the first class, and allows the utility function parameters for these classes to then also reflect this same divide.

\section{Results}
Tables of parameter estimates for utility functions and class membership probabilities are presented in the appendix. Perhaps more interesting for interpretation are the marginal effects of the various variables impacting these utility functions. Given the nonlinearity of our model, the calculation of these marginals is not as straight-forward as simply interpreting a slope coefficient.  Marginal effects for the parameters of interest are computed using automatic differentiation software. I follow Cameron \& Trivedi (2005) in defining the relevant marginal effects as ~$\frac{1}{N}\sum_i \frac{\partial P_{ij}(m)}{\partial \hat X_i}$. This is the average change in choice probability across all observations with changes in the given ~$x$ variable. Note that across the five modalities, the marginal effects of the same variable will sum to zero\footnote{It is this ``zero sum" characteristic that allows us to speak to substitution between these marriage modalities.} and that due to the interrelationships between marriage modality choices, probabilities of a given choice will have marginal effects with respect to variables that do not directly enter their utility function.

\begin{table}[H]
	\centering
	\resizebox{.9\textwidth}{!}{%
		\begin{tabular}{@{\extracolsep{5pt}}lccccc}
			\hline
			\hline
			& \multicolumn{5}{c}{\textit{Marriage Modality}} \\
			\cline{2-6}
			\\[-1.8ex] & Love Marriage & Mock Kidnapping & Arranged Marriage & Bride Capture & Forgo \\
			\hline \\[-1.8ex]
			Aksakal Governance & -0.0451 & -0.0211 & 0.0327 & 0.0165 & 0.0178 \\
			Police &-0.0197 & -0.0001 &-0.0229 & 0.0550 & -0.0130 \\
			Kalym Price & 0.0003 & -0.0050 & -0.0082 & 0.0082 & 0.0049 \\
			Household Income & -0.0157 & -0.0003 & 0.0116 & 0.0023 & 0.0023 \\
			Second Home Ownership & -0.1260 & -0.0359 & 0.1128 & 0.0595 & -0.0089 \\
			Vehicle Ownership & 0.0682 & 0.0101 & -0.0064 & -0.0699 & -0.0023 \\
			Loan Availability & 0.0835 & 0.0280 & 0.0023 & -0.0627 & -0.0540 \\
			Event Host & 0.0323 & -0.0268 & 0.0143 & -0.0336 & 0.0141 \\
			Employed & -0.1261 & -0.0296 & 0.2811 & -0.0179 & -0.1118 \\
			\hline
			Base Probability & 0.2015 & 0.0272 & 0.3482 & 0.1847 & 0.2383\\
			\hline
			\hline \\[-1.8ex]
		\end{tabular}
	}
	\caption{Table depicting the marginal impacts of explanatory variables on the probability of a man selecting into a given marriage modality- across both classes.}
\end{table}

\section{Discussion}
The marginal effects table presented in preceding section allows us to examine patterns of substitution between marriage modalities. Several interesting features stand out here. First, across multiple metrics it appears that lower status men are pushed towards elopement/mock kidnappings, non-consensual kidnappings, or forgoing marriage. Employment status seems to be the largest measure of a man's social status in regards to the marriage market. Employed men are much more likely to have an arranged marriage, substituting away from love marriages primarily. Men who are of lower status across almost all of our proxies are much more likely to kidnap a bride or forgo marriage entirely. Similarly, increases in \emph{kalym} prices lead to an increased probability of non-consensual kidnapping or of forgoing marriage. Men may risk being priced out of marriage, and in turn face a choice between kidnapping or being alone. That police presence is positively related to the probability of bride capture may be due to two features. As documented in Sheilds (2006) police are highly unlikely to take complaints of bride kidnapping seriously. This may in turn reduce the deterrent effect of their presence. Further, there is a police presence across most of the communities in our data. The lack of variability across this measure could also be a source of bias.

Perhaps somewhat surprisingly, running against the marginal effects of the other status proxies, second home ownership and household income increase the probability of bride capture. This may be because, as in historic times, bride capture can lead to significant disputes between families. As discussed in Aktas (2021), the disputes these incidents can spur ensured that only wealthy families who could afford such a conflict obtained brides through capture in historic times. This seems to point to a difference between wealth and social status present within Kyrgyz communities.

\emph{Aksakal} governance makes men more likely to engage in bride capture. Despite being a relatively small 1.65 percentage point increase in the probability of an arbitrary man engaging in \emph{ala kachuu}, this corresponds to a 9\% increase over the baseline probability. Further, for men of traditional value sets this is a 3.06 percentage point increase. Perhaps more alarmingly, \emph{aksakal} governance corresponds with a 1.10 percentage point increase in the probability of kidnapping even for men who in their surveys espoused very modern and egalitarian values. It seems that living in a community that is governed by institutions that validate violence against women leads more men to elect to engage in such violence, even if these men outwardly support gender equality. Governance by \emph{aksakal} drastically changes the dynamics of marriage within a community. Men substitute away from love matches and instead see increases in the probability of kidnapping, arranged marriages or remaining alone. 

It appears that throughout these results, low status men are choosing between kidnapping and what I have empirically defined here as the outside option: forgoing marriage entirely. This is consistent with what is discussed in Werner et al. (2018): kidnappers are largely low status men who have been rejected in the past. These men are possibly priced out of the consensual marriage market. For each increase in local \emph{kalym} prices equal in value to one sheep, men grow 2.1\% more likely to forgo marriage and 4.4\% more likely to capture a bride. This speaks in some regards to Leeson \& Suarez (2017). Men who are unable to find wives through more socially acceptable means, unfortunately often pursue more fringe alternative modalities. This also analogues patterns regarding the perpetrators of contemporary violence against women. Psychological analysis of sexual offenders has found that male perpetrators often espouse a general lack of intimacy through their life (Marshall 1989). This is indicative of their inability or failure to find partners through socially normalized routes. When looked at alongside the result regarding \emph{aksakal} governance, this suggests that across societies socially alienated individuals who do not perceive a significant risk of punishment are at a greater probability of engaging in violence against women.

In this paper, I have explored the factors influencing marriage modality choice in Kyrgyzstan with a particular emphasis on the role played by governance by informal institutions. It appears that since men governed by traditional elders, rather than the state, have little to fear in terms of institutional punishments for kidnapping they are more likely to find \emph{ala kachuu} a legitimate means of finding a wife. By estimating a model of discrete choice, I have been able to analyze patterns of substitution in marriage modality choice, documenting evidence that suggests that men on the fringes of society are largely choosing between kidnapping and forgoing marriage all together.

\clearpage

\section{Statements and Declarations}

\textbf{Data and Code Availability}: The datasets analyzed during the current study are available through application to IZA, [\url{https://datasets.iza.org//dataset/124/life-in-kyrgyzstan-panel-study-2013}]. Analysis was conducted using the R and Julia programming languages. Code is available from the author upon request.

\medskip

\noindent\textbf{Competing Interests}: The author has no competing interests to declare that are relevant to the content of this article.

\medskip

\noindent\textbf{Funding}: The author did not receive support from any organization for the submitted work.

\appendix
\section{Appendix}

\subsection{Table of All Utility Function Parameters}
\begin{table}[H]
	\centering
	\resizebox{\textwidth}{!}{%
		\begin{tabular}{@{\extracolsep{4pt}}lcccccccccc}
			\hline
			\hline
			& \multicolumn{10}{c}{\textit{Marriage Modality}} \\
			\cline{2-11}
			& \multicolumn{2}{c}{Love Marriage} & \multicolumn{2}{c}{Mock Kidnapping} & \multicolumn{2}{c}{Arranged Marriage} & \multicolumn{2}{c}{Bride Capture} & \multicolumn{2}{c}{Choice Match Nest} \\
			\cline{2-3} \cline{4-5} \cline{6-7} \cline{8-9} \cline{10-11}
			& Class 1 & Class 2 & Class 1 & Class 2 & Class 1 & Class 2 & Class 1 & Class 2 & Class 1 & Class 2 \\
			\hline
			Intercept & -0.273 & -0.060 & 0.536 & -9.289 & 0.067 & -2.160 & 0.106 & -0.705 & -5.338 & 0.460 \\
			& (0.0000) & (0.0003) & (0.0004) & (0.0104) & (0.0002) & (0.0012) & (0.0001) & (0.0003) & (0.0066) & (0.0003) \\
			Aksakal Governance & - & - & -0.367 & -14.932 & - & - & 0.038 & -0.025 & -0.271 & -0.254 \\
			& - & - & (0.1048) & (0.0000) & - & - & (0.0940) & (0.0856) & (0.0663) & (0.0052) \\
			Police Presence & - & - & - & - & - & - & 0.004 & 0.353 & - & - \\
			& - & - & - & - & - & - & (0.0001) & (0.0538) & - & - \\
			Kalym Price & -0.005 & 0.051 & -0.087 & 0.288 & 0.003 & -0.054 & -0.036 & 0.076 & -0.692 & -0.007 \\
			& (0.0001) & (0.0006) & (0.0000) & (0.0011) & (0.0001) & (0.0006) & (0.0001) & (0.0005) & (0.0012) & (0.0005) \\
			Household Income & -0.006 & 0.009 & 0.002 & 1.067 & 0.005 & 0.019 & -0.000 & 0.009 & -0.151 & -0.079 \\
			& (0.0016) & (0.0005) & (0.0006) & (0.0012) & (0.0005) & (0.0005) & (0.0000) & (0.0005) & (0.0002) & (0.0005) \\
			Second Home Ownership & -2.584 & 0.855 & 0.045 & -10.003 & 0.904 & -0.630 & 0.261 & 0.088 & 2.969 & -0.463 \\
			& (0.0016) & (0.0003) & (0.0003) & (0.0075) & (0.0008) & (0.0003) & (0.0007) & (0.0003) & (0.0037) & (0.0003) \\
			Vehicle Ownership & -0.009 & -1.477 & 0.013 & -2.820 & -0.171 & 0.282 & -0.000 & -0.440 & 2.244 & 0.922 \\
			& (0.0000) & (0.0008) & (0.0004) & (0.0033) & (0.0002) & (0.0003) & (0.0001) & (0.0003) & (0.0028) & (0.0003) \\
			Loan Availability & 0.326 & -0.244 & 0.101 & -10.004 & -0.212 & 0.676 & 0.003 & -0.190 & 6.904 & 0.606 \\
			& (0.0000) & (0.0003) & (0.0004) & (0.0075) & (0.0002) & (0.0003) & (0.0001) & (0.0003) & (0.0085) & (0.0003) \\
			Event Host & 0.014 & -0.077 & -1.055 & -10.002 & -0.007 & -0.000 & -0.046 & -0.168 & 8.382 & 0.070 \\
			& (0.0000) & (0.0003) & (0.0007) & (0.0075) & (0.0002) & (0.0003) & (0.0001) & (0.0003) & (0.0103) & (0.0003) \\
			Employed & -0.019 & -0.309 & -0.007 & -10.479 & 0.041 & 2.024 & -0.002 & 0.183 & -9.951 & 0.565 \\
			& (0.0000) & (0.1826) & (0.0004) & (0.1718) & (0.0002) & (0.0404) & (0.0001) & (0.0589) & (0.0122) & (0.0966) \\
			\hline
			Observations & \multicolumn{10}{c}{3340} \\
			BIC & \multicolumn{10}{c}{-8698.22} \\
			\hline
			\hline
			\textit{Note:} & \multicolumn{10}{r}{Standard errors are computed with the robust estimator of Greene (2018).} \\
			\textit{} & \multicolumn{10}{r}{Utility for the outside option, forgoing marriage, is normalized to 0.} \\
			\hline
		\end{tabular}
	}
	\caption{Table depicting all estimated utility function parameters with standard errors in parentheses.}
\end{table}

\subsection{Table of Parameters for Class Membership}
\begin{table}[H]
	\centering
	\resizebox{.7\textwidth}{!}{%
		\begin{tabular}{@{\extracolsep{4pt}}lcc}
			\hline
			\hline
			& \multicolumn{2}{c}{\textit{Class Membership Probabilities}} \\
			\cline{2-3}
			& Class 1 & Class 2 \\
			\hline
			Intercept & 0.1848 & -0.1086 \\
			& (0.5008) & (0.5008) \\
			Kyrgyz & 1.1524 & -0.0053 \\
			& (0.0924) & (0.0924) \\
			Village Resident & 0.0460 & -0.7944 \\
			& (0.0398) & (0.0399) \\
			Husband Choice Priority & 0.0488 & -0.2158 \\
			& (0.1427) & (0.1426) \\
			Woman Belongs at Home & 2.3454e-9 & -8.8376e-10 \\
			& (0.0004) & (0.0004) \\
			Woman Only Fulfilled as Mother & 7.5175e-10 & -2.9502e-10 \\
			& (0.0872) & (0.0871) \\
			Husband's Career is More Important & 2.1579e-9 & -5.5188e-10 \\
			& (0.1090) & (0.1090) \\
			Male Education Importance & 7.2491e-9 & -2.0413e-9 \\
			& (0.1088) & (0.1088) \\
			Fulfillment as Housewife & 1.0651e-9 & -6.1535e-10 \\
			& (0.4050) & (0.4049) \\
			A Wife Should Not Work & 0.0981 & -0.0358 \\
			& (0.0482) & (0.0482) \\
			A Wife Should Not Complain & 1.4789e-10 & -6.3851e-10 \\
			& (0.5710) & (0.5709) \\
			A Wife Should be Obedient & 6.6794e-7 & -2.0419e-7 \\
			& (0.2410) & (0.2411) \\
			School in Community & 0.0115 & -0.8284 \\
			& (0.4862) & (0.4863) \\
			Can Rely on Others & 0.0678 & -0.0145 \\
			& (0.5008) & (0.5008) \\
			Women Can Enjoy Work & -2.5267e-10 & 3.1770e-9 \\
			& (0.7472) & (0.7472) \\
			Dual Income is Important & -3.4054e-10 & 1.4201e-9 \\
			& (0.3920) & (0.3919) \\
			A Wife Should be Confident & -0.3480 & -0.0255 \\
			& (0.5918) & (0.5918) \\
			A Wife Should be Respected by Community & 0.2585 & -0.4601 \\
			& (0.0216) & (0.0215) \\
			General Trust & 0.5855 & 0.3372 \\
			& (0.0610) & (0.0610) \\
			Cannot Rely on Others & -0.4877 & -0.0916 \\
			& (0.0581) & (0.0581) \\
			Caution is Needed in the Community & -2.5570 & 0.2703 \\
			& (0.7158) & (0.7170) \\
			Trust in Family & -0.0440 & 0.4111 \\
			& (0.2852) & (0.2851) \\
			Trust in Neighbors & -0.1601 & -0.1945 \\
			& (0.1805) & (0.1806) \\
			Trust in Strangers & 0.0018 & 0.0321 \\
			& (0.1135) & (0.1134) \\
			Trust other Ethnicities & 0.0439 & -0.0219 \\
			& (0.1448) & (0.1448) \\
			\hline
			\hline
			\textit{Note:} & \multicolumn{2}{r}{Standard errors are computed with the robust estimator.} \\
			\hline
		\end{tabular}
	}
	\caption{Estimated parameters for class membership probabilities with respect to various social and personal attitudes. Details regarding survey questions and coding are in class estimation section of paper.}
\end{table}

\end{document}